# Symmetry relations revealed in Mueller Matrix hemispherical maps

Adrian J. Brown[*,1], Yu Xie[2]

[1] SETI Institute, 189 Bernardo Ave, Mountain View, CA 94043, USA
[2] Department of Atmospheric Sciences, Texas A&M University, College Station TX 77843, USA

[*] Corresponding author address: Adrian J. Brown, SETI Institute, 189 Bernardo Ave, Mountain View, CA 94043; Tel: +1 (650) 810-0223; email: abrown@seti.org

# HIGHLIGHTS

- Hemispherical Mueller matrix maps for spheres and hexagonal cylinders with zenith and off-zenith illumination are studied.
- Symmetry relationships for Mueller Matrix hemispherical observations for zenith and off-zenith illumination are identified.
- Limits of single-scattering Mueller matrix symmetry relationships are shown based on numerical simulations.

## Abstract

The hemispherical Mueller matrix map for light reflected from a plane-parallel planetary atmosphere is shown to obey several symmetry properties that provide a straightforward method to check their physical realizability. The mirror scattering relation and the reciprocity relation are employed to support the symmetry rules for hemispherical Mueller matrix maps. Additionally, two classes of experiments are identified in which the symmetry rules can be applied, namely, when the incident beam is at zenith and off-zenith angles and the scattering particles are spheres and randomly oriented hexagonal particles.

# 1. Introduction

In this study, we explicitly identify symmetry rules in reflected hemispherical Mueller matrix maps. Previous research into the light scattering by ice crystals [1] has incorrectly shown results that violated these rules. We present corrected hemispherical maps and provide some simple rules to use for a straightforward accuracy check of hemispherical Mueller matrix maps from theoretical and numerical perspectives.

# 2. Mirror symmetry and reciprocity relations

The intensity and polarization state of a beam of light can be specified in terms of the Stokes vector by [2, 3]

$$\mathbf{I} = \begin{pmatrix} I \\ Q \\ U \\ V \end{pmatrix} = \begin{pmatrix} <E_\parallel E_\parallel^*> + <E_\perp E_\perp^*> \\ <E_\parallel E_\parallel^*> - <E_\perp E_\perp^*> \\ <E_\parallel E_\perp^*> + <E_\perp E_\parallel^*> \\ i(<E_\parallel E_\perp^*> - <E_\perp E_\parallel^*>) \end{pmatrix} = \begin{pmatrix} a^2 \\ a^2 \cos 2\beta \cos 2\chi \\ a^2 \cos 2\beta \sin 2\chi \\ a^2 \sin 2\beta \end{pmatrix}, \quad (1)$$

where $E_\parallel$ and $E_\perp$ are the respective electric vectors parallel and perpendicular to a chosen reference plane, $a$ is the distance between endpoints for the major and minor axes of the polarization ellipse, $\tan\beta$ is the ellipticity (the ratio of minor to major axis of polarization ellipse), and $\chi$ is the orientation angle of the polarization ellipse.

The Mueller matrix, $M$, transforms an input Stokes vector, $I_0$, to the output Stokes vector, $I$, associated with a certain physical process (e.g., the scattering of the incident radiation by a particle):

$$\mathbf{I} = M\mathbf{I}_0 = \begin{pmatrix} M_{11} & M_{12} & M_{13} & M_{14} \\ M_{21} & M_{22} & M_{23} & M_{24} \\ M_{31} & M_{32} & M_{33} & M_{34} \\ M_{41} & M_{42} & M_{43} & M_{44} \end{pmatrix} \begin{pmatrix} I_0 \\ Q_0 \\ U_0 \\ V_0 \end{pmatrix}. \quad (2)$$

To study the symmetry relations of the Mueller matrix elements, we begin with the reciprocity and mirror symmetry relations. Interested readers are referred to van de Hulst [2], Hovenier [4], and Hovenier et al.[5] for in-depth discussions of the those relations.

The physical law of reciprocity implies that

$$F = \Delta_3 F^T \Delta_3, \quad (3)$$

where $^T$ indicates transpose and

$$\Delta_3 = \begin{pmatrix} 1 & 0 & 0 & 0 \\ 0 & 1 & 0 & 0 \\ 0 & 0 & -1 & 0 \\ 0 & 0 & 0 & 1 \end{pmatrix}. \qquad (4)$$

For a target with mirror symmetry, the following relationship holds:

$$F(\theta) = \Delta_{3,4} F(\theta) \Delta_{3,4}, \qquad (5)$$

where

$$\Delta_{3,4} = \begin{pmatrix} 1 & 0 & 0 & 0 \\ 0 & 1 & 0 & 0 \\ 0 & 0 & -1 & 0 \\ 0 & 0 & 0 & -1 \end{pmatrix}. \qquad (6)$$

Therefore, the scattering matrix, F, governed by the laws of reciprocity and mirror symmetry has a reduced number of independent elements given by (van de Hulst [2] page 50)

$$F(\theta) = \begin{bmatrix} a_1(\theta) & b_1(\theta) & 0 & 0 \\ b_1(\theta) & a_2(\theta) & 0 & 0 \\ 0 & 0 & a_3(\theta) & b_2(\theta) \\ 0 & 0 & -b_2(\theta) & a_4(\theta) \end{bmatrix}. \qquad (7)$$

This relationship is valid for spheres, randomly oriented cylinders and ellipsoids, and randomly oriented nonspherical particles (e.g., hexagonal particles) with mirror images in equal numbers. In the current study, we re-capitulate the basic method of deriving the rotations required for perpendicular illumination given by Hovenier and de Haan [6], but with a slightly simplified nomenclature.

When a beam of light is rotated counterclockwise (looking in the direction of propagation) through an angle $\alpha$, the rotation matrix, $L(\alpha)$, relating initial and final Stokes vectors is (e.g., Hovenier [5] Eq. 1.51)

$$L(\alpha) = \begin{pmatrix} 1 & 0 & 0 & 0 \\ 0 & \cos 2\alpha & \sin 2\alpha & 0 \\ 0 & -\sin 2\alpha & \cos 2\alpha & 0 \\ 0 & 0 & 0 & 1 \end{pmatrix}. \qquad (8)$$

Consider a beam of arbitrarily chosen incident light with a reference azimuth plane, $\phi_0$, (see Figure 1a). To derive the light scattered through another reference plane, $\phi$, the Stokes vector referenced to plane $\phi_0$ must be rotated through an angle $\phi-\phi_0$. Thus,

$$F(\mu_0 = 1, \phi - \phi_0) = F(\mu_0 = 1, \phi_0 = 0)L(\phi - \phi_0). \tag{9}$$

Therefore, using Eqs. (7) and (8) to find the Mueller matrix for perpendicularly incident light, which is initially along plane $\phi_0$ and is subsequently reflected along plane $\phi$, the following equation is obtained (e.g., Hovenier [5] Eq. 4.106):

$$F(\mu_0 = 1, \phi - \phi_0) = \begin{pmatrix} a_1 & b_1 \cos 2(\phi - \phi_0) & b_1 \sin 2(\phi - \phi_0) & 0 \\ b_1 & a_2 \cos 2(\phi - \phi_0) & a_2 \sin 2(\phi - \phi_0) & 0 \\ 0 & -a_3 \sin 2(\phi - \phi_0) & a_3 \cos 2(\phi - \phi_0) & b_2 \\ 0 & b_2 \sin 2(\phi - \phi_0) & -b_2 \cos 2(\phi - \phi_0) & a_4 \end{pmatrix}. \tag{10}$$

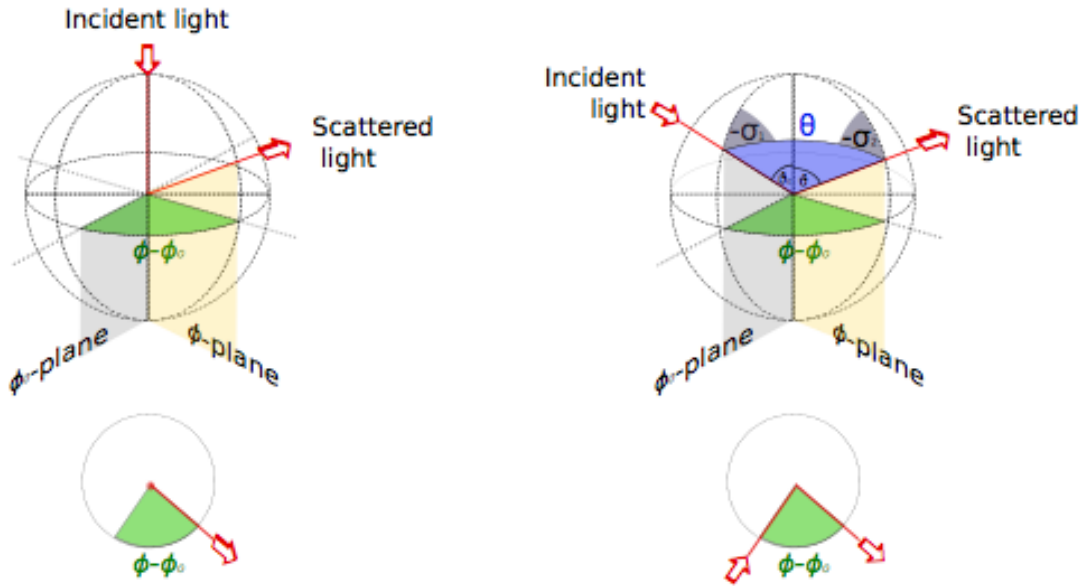

Figure 1. a.) Illustration of rotation of the incident light from the zenith, referenced to the $\phi_0$-plane (plane of incidence), into the $\phi$–plane (scattering plane) when rotated by an angle $\phi$-$\phi_0$ and multiplied by the rotation matrix L($\phi$-$\phi_0$). b.) Same diagram for illumination from off-zenith, referenced to $\phi_0$-plane (plane of incidence), into the $\phi$–plane (scattering plane) when prerotated by an angle -$\sigma_1$ and postrotated by an angle -$\sigma_2$.

For off-zenith illumination (Figure 1b), two rotations of the plane of reference are required to translate the incident beam into the scattered beam, amounting to a premultiplication by L(-$\sigma_1$) and postmultiplication by L($\pi$-$\sigma_2$) or, equivalently, L(-$\sigma_2$):

$$F(\vartheta,\varphi,\vartheta',\varphi') = L(\pi - \sigma_2)F(\theta)L(-\sigma_1) = L(-\sigma_2)F(\theta)L(-\sigma_1). \tag{11}$$

Using Eqs. (7) and (8) to complete this multiplication gives (e.g., Hovenier [5] Eq. 3.9)

$$F(\mu,\mu_0,\phi - \phi_0) =$$
$$\begin{pmatrix} a_1 & b_1\cos 2\sigma_1 & -b_1\sin 2\sigma_1 & 0 \\ b_1\cos 2\sigma_2 & \cos 2\sigma_2 a_2 \cos 2\sigma_1 - \sin 2\sigma_2 a_3 \sin 2\sigma_1 & -\cos 2\sigma_2 a_2 \sin 2\sigma_1 - \sin 2\sigma_2 a_3 \cos 2\sigma_1 & -b_2\sin 2\sigma_2 \\ b_1\sin 2\sigma_2 & \sin 2\sigma_2 a_2 \cos 2\sigma_1 + \cos 2\sigma_2 a_3 \sin 2\sigma_1 & -\sin 2\sigma_2 a_2 \sin 2\sigma_1 + \cos 2\sigma_2 a_3 \cos 2\sigma_1 & b_2\cos 2\sigma_2 \\ 0 & -b_2\sin 2\sigma_1 & -b_2\cos 2\sigma_1 & a_4 \end{pmatrix} \tag{12}$$

2.1 Relationships for Backscattering direction, $\theta=\pi$ for zenith illumination

As first pointed out by van de Hulst [2] and discussed in Mishchenko and Hovenier [7] and Hovenier and Mackowski [8], for single scattering from a randomly oriented, rotationally symmetric particle or an ensemble of particles with a plane of symmetry in the $\theta=\pi$ (backscattering) direction, the following Mueller matrix holds (e.g., Hovenier [5] Eq. 2.73):

$$F = \begin{bmatrix} a_1 & 0 & 0 & 0 \\ 0 & a_2 & 0 & 0 \\ 0 & 0 & -a_2 & 0 \\ 0 & 0 & 0 & a_4 \end{bmatrix} \tag{13}$$

Furthermore, by using the reciprocity principle it is possible to show that (e.g., Hovenier [5] Eq. 2.76)

$$a_4 = a_1 - 2a_2 \tag{14}$$

This provides a powerful restriction on the allowed values of elements $M_{11}$, $M_{22}$, $M_{33,}$ and $M_{44}$ for scattering in the $\theta=\pi$ direction (i.e., in the central section of the hemispherical maps). For example, as shown in Hovenier and Mackowski [8], $M_{11} \geq M_{22} \geq 0$ because $M_{11} \geq M_{44}$ [9].

According to Hovenier and Mackowski [8], Table 2, the following relationship also holds for single scattering from targets consisting of homogenous, nonactive spheres:

$$a_1 = a_2 = -a_4 \tag{15}$$

this will be discussed further in the numerical results section.

2.2 Relationships for Backscattering direction, $\theta=\pi$ for off-zenith illumination

To compute the transfer matrix for the backscattering direction for off-zenith illumination, Eq. (12) can be used. Let $\sigma_1$ approach 0 (letting the scattering axis in Figure 1b approach zenith) in order for $\cos(2\sigma_1)=1$ and $\sin(2\sigma_1)=0$. This leads to the following relation (e.g., Hovenier [5] Eq. 4.113):

$$F(\mu_0, \mu=1, \sigma_2 = \phi - \phi_0) = \begin{pmatrix} a_1 & b_1 & 0 & 0 \\ b_1 \cos 2\sigma_2 & a_2 \cos 2\sigma_2 & -a_3 \sin 2\sigma_2 & -b_2 \sin 2\sigma_2 \\ b_1 \sin 2\sigma_2 & a_2 \sin 2\sigma_2 & a_3 \cos 2\sigma_2 & b_2 \cos 2\sigma_2 \\ 0 & 0 & -b_2 & a_4 \end{pmatrix} \quad (16)$$

## 3. Numerical experiments: hemispherical Mueller matrix maps

The appearance of hemispherical Mueller matrix elements, illuminated from the zenith or off-zenith and projected into a 2-dimensional map, displays readily apparent patterns that can be used as a straightforward accuracy check. The results from two numerical experiments exercising the theory are presented.

3.1 Experiment I: Zenith (perpendicular) illumination

Figure 2 shows the hemispherical Mueller maps created using the adding-doubling code [10] coupled with a Mie scattering code to expand the scattering matrix in generalized spherical coordinates. The model parameters are: $\mu_0=1$; optical depth $\tau=1$; wavelength $\lambda=0.7$; and, the target atmosphere used one homogenous layer with no reflecting sub-surface boundary (i.e. semi-infinite conditions). The layer consisted of spheres with an optical index n=1.33, k=0.0 and with a size distribution given by the modified gamma distribution as parameterized in de Rooij [11]:

$$n(r) = \frac{\gamma}{r_c \Gamma\left(\frac{\alpha+1}{\gamma}\right)} \left(\frac{\alpha}{\gamma}\right)^{\frac{\alpha+1}{\gamma}} \left(\frac{r}{r_c}\right)^\alpha \cdot \exp\left[-\frac{\alpha}{\gamma}\left(\frac{r}{r_c}\right)^\gamma\right] \quad (17)$$

where the mode radius is $r_c=0.07\mu m$, $\alpha=2$, and $\gamma=0.5$. The size distribution is equivalent to the water haze L of Deirmendijian [12]).

The top right and bottom left 1x2 blocks in the single scattering matrix are zero (see Eq. (10)), which makes the $M_{14}$, $M_{24}$, $M_{31}$, and $M_{41}$ elements of the Mueller Matrix uniformly zero.

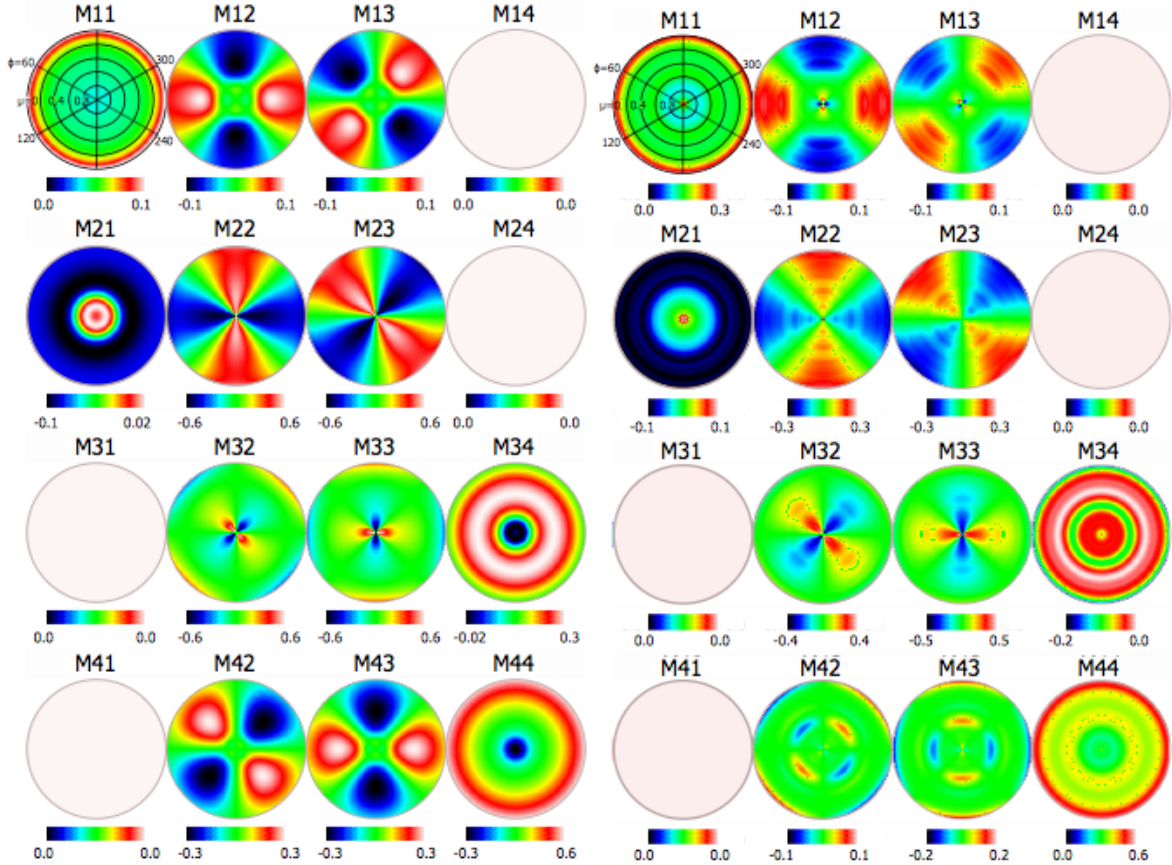

Figure 2. Mueller matrix hemispherical map for zenith scattering of a.) Mie spheres equivalent to Deirmendijian (1969) L water haze and b.) Hexagonal cylinder particles with 2a/L=80μm/300μm. See text for model setup. All elements have been normalized with the $M_{11}(\theta,\phi)$ element at the same $\theta,\phi$ coordinate.

The maps for elements $M_{11}$, $M_{21}$, $M_{34}$, and $M_{44}$ project into 'circles'; whereas, elements $M_{*2}$, and $M_{*3}$ (i.e., the central two columns) display 'crosses'.

The circles appear because the $M_{11}$, $M_{21}$, $M_{34}$, and $M_{44}$ elements have no azimuth dependence, as shown in Eq. (10).

Although it is difficult to observe the values of $M_{11}$, $M_{22}$, and $M_{44}$ in the central part of the Mueller Matrix maps in Figure 2, the relationships in Eqs. (12) and (13) hold, at least for small optical depths ($\tau$=0.1). For scattering from Deirmendijian's water haze L with $\tau$=0.1, the following Mueller matrix results for the $\theta=\pi$ direction (closely related to the central point of the 16 images in Figure 2; however, the optical depth is increased ($\tau$=1) in Figure 2):

$$F = \begin{bmatrix} 0.0025 & 0 & 0 & 0 \\ 0 & 0.0024 & 0 & 0 \\ 0 & 0 & -0.0024 & 0 \\ 0 & 0 & 0 & -0.0023 \end{bmatrix} \quad (18)$$

The Mueller matrix in Eq. (18) is clearly not quite compliant with the relationship in Eq. (15) as a result of the breakdown of the single scattering assumptions inherent in constructing the latter.

A close relationship exists between elements $M_{12} \rightarrow M_{13}$, $M_{22} \rightarrow M_{23}$, $M_{32} \rightarrow M_{33}$, and $M_{42} \rightarrow M_{43}$ as they are linearly related after rotations of 45° in the counterclockwise direction (by convention, this is positive rotation and is noted in Hovenier [5] p.11).

This rotational relationship can be understood by observing the exchange of cos and sin functions in Eq. (10).

3.2 Experiment II: Off-zenith (non-perpendicular) illumination

Lawless et al. [1] presented off-zenith hemispherical Mueller Matrix maps, but unfortunately, the maps were incorrectly calculated.

Lawless et al. [1] used an adding-doubling vector radiative transfer code [10] to calculate the Mueller Matrix hemispherical scattering maps of hexagonal columns and plates. Figure 3 (shaded) shows the Lawless et al. results for off-zenith (30° incidence angle) illumination of the two target types. The maps were prepared by calculating the backscattered Mueller Matrices for phase angles of 0-180° and mirroring these results in the 180-360° phase angle quadrant. This unfortunately led to incorrect results, which can be verified with the symmetry relations in Eqs. (10) to (16).

Figure 3 also shows the corrected version for the hexagonal cylinders and plates. The target parameters were: $\mu_0=0.846$ (30° incidence angle); optical depth $\tau=1$; wavelength $\lambda=0.66$; and, the target atmosphere used one homogenous layer with no reflecting sub-surface. The layer consisted of identical hexagonal particles with aspect ratios of $2a/L=80\mu m/300\mu m$ (cylinders) and $2a/L=300\mu m/80\mu m$ (plates), where L and a are the length and radius of a single hexagonal particle. The optical index is n=1.3078, k=1.66x10$^{-8}$.

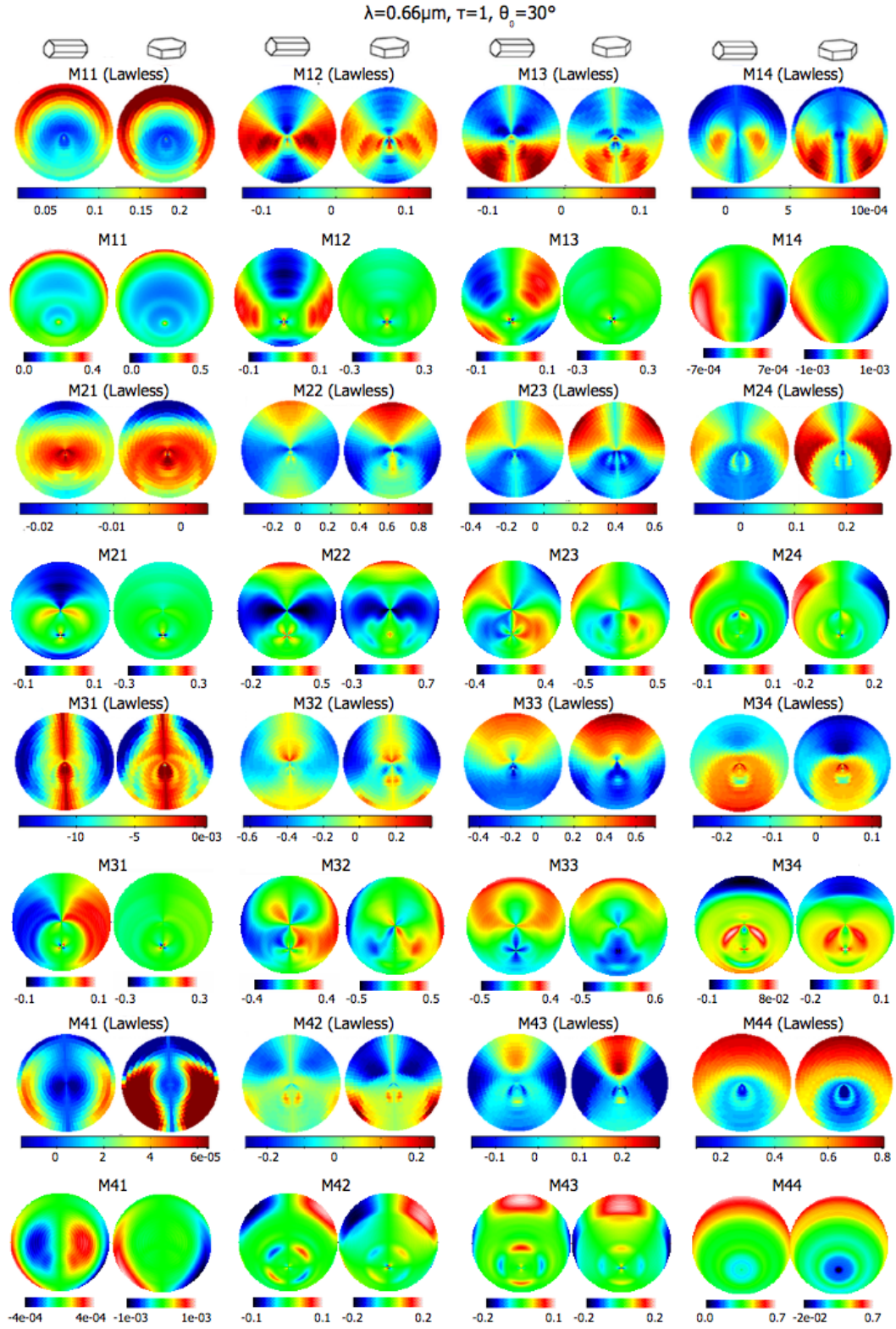

Figure 3. Figure 6 from Lawless et al. [1] (shaded), mixed with corrected hemispherical Mueller maps from this study for hexagonal cylinders (left) and plates (right). All elements have been normalized with the $M_{11}(\theta,\phi)$ element at the same $\theta,\phi$ coordinate.

Figure 4 shows an example of an off-zenith observation of Mie scatterers. The same target is used as for Figure 2: optical depth $\tau=1$; wavelength $\lambda=0.7$; and, one homogenous layer with no reflecting sub-surface. The layer consisted of spheres with optical index n=1.33, k=0.0 and a size distribution matching the water haze L of Deirmendijian [12]. The incidence angle, $\theta_0$, is 30° (same as for the hexagonal particles in Figure 3).

When the illumination source is moved to an off-zenith (non-perpendicular) orientation, the symmetry of the resultant hemispherical Mueller matrix maps is decreased; however, most patterns remain and we relate these to our symmetry equations.

The 'crosses' present in the zenith illumination case shift orientation according to the new incidence angle in off-zenith orientation. The 'odd-even' behavior of Eq. (10) is repeated in Eq. (12) for the off-zenith case – the top right and bottom left quadrants display anti-symmetry in the plane of incidence (also at 90° to this plane – the x-y axis in the figures), and the top left and bottom right quadrants display mirror symmetry. The results are due to the positions of odd (sin, sin.cos) and even (cos, sin.sin, cos.cos) elements in Eq. (12).

The 'circles' for $M_{11}$ and $M_{44}$ remain, but are slightly offset from the center of the hemispherical maps due to the increased inclination angle. Elements $M_{*2}$ and $M_{*3}$ (i.e., the central two columns) display 'crosses' for both hexagonal particles and spheres, although these have become somewhat distorted. Elements $M_{21}$ and $M_{34}$ no longer display circles, but, instead their shape becomes carotoid for spheres.

There is a close relationship between the $M_{12} \rightarrow M_{13}$, $M_{22} \rightarrow M_{23}$, $M_{32} \rightarrow M_{33}$, and $M_{42} \rightarrow M_{43}$ elements and they are linearly related after 45° rotations. The elements in columns from $M_{*2}$ to the element in the next column, $M_{*3}$, have each rotated counterclockwise 45°, when looking in the direction of propagation.

The reason for this relationship can be grasped by observing the exchange of cos and sin trigonometric functions in Eq. (12). In the off-zenith case, the relationship is somewhat obscured because of the more complex shapes in elements $M_{23}$ and $M_{33}$; however, the 45° relationship continues to exist for spheres and hexagonal targets.

Elements $M_{41}$ and $M_{14}$ have maxima and minima very close to zero (both elements are less than $8\times10^{-4}$). According to Eq. (12), which reflects the off-zenith illumination situation, these elements should be completely null. Again, this small deviation is likely due to the breakdown of the single scattering assumptions inherent in constructing Eq. (12).

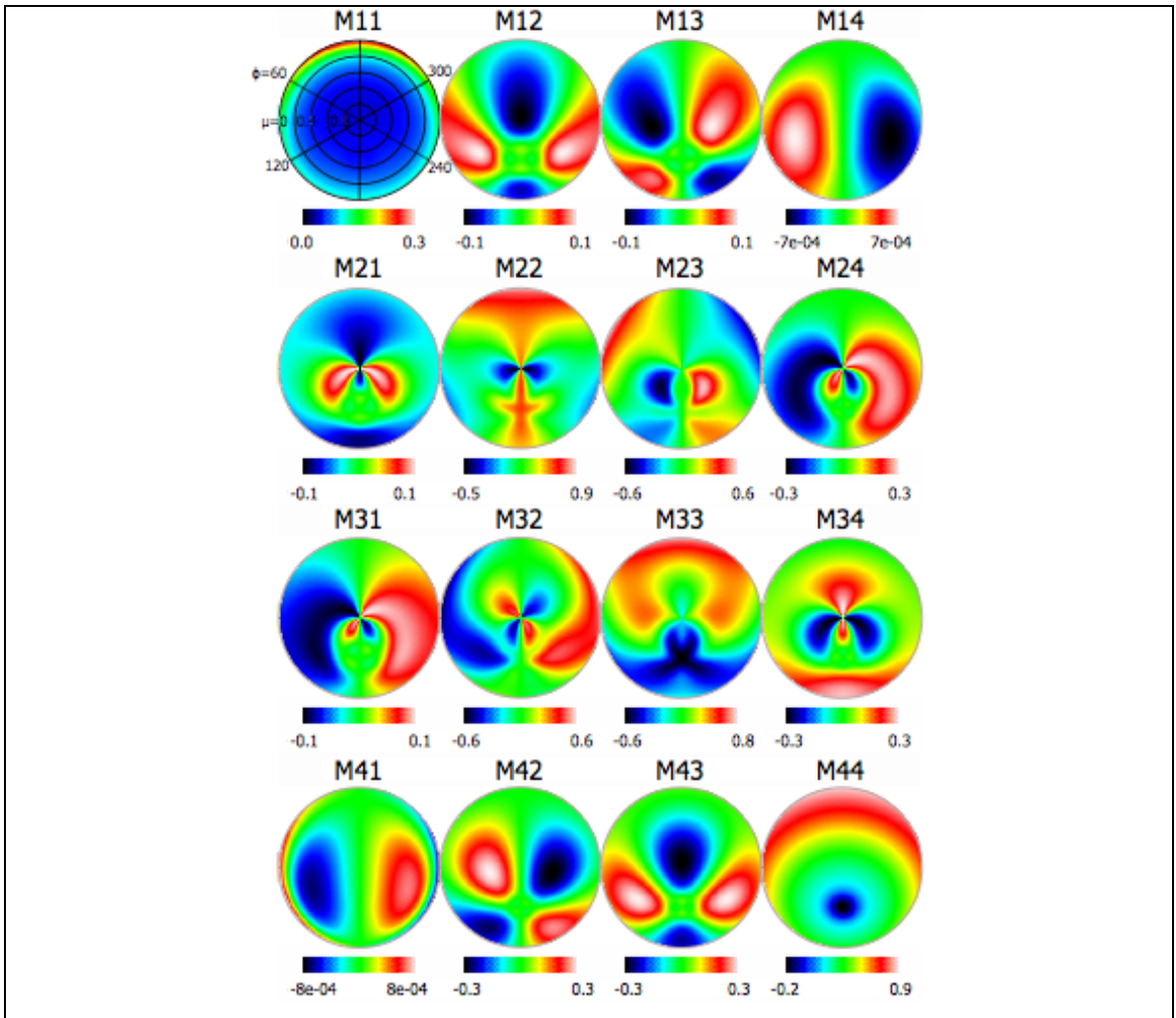

Figure 4. Off-zenith Mueller matrix hemispherical map for scattering from Mie spheres equivalent to Deirmendijian [12] L water haze. See text for model setup. All elements have been normalized with the $M_{11}(\theta,\phi)$ element at the same $\theta,\phi$ coordinate.

For the zenith illumination, the scattering into the backscattering direction ($\theta=\pi$) displays a special relationship described by Eq. (16). Because this is difficult to decipher in the maps, the matrix is reproduced for the spherical target whose hemispherical maps are shown in Figure 4. The angle is $\phi=30°$ ($\theta_0=60°$, $\theta=\pi$) and the optical depth is lower ($\tau=0.1$) in order to reduce multiple scattering effects (as discussed in section 2.1.3). The result is

$$F = \begin{bmatrix} 0.0043 & -0.0008 & 0 & 0 \\ -0.0004 & 0.0019 & -0.0012 & -0.0016 \\ -0.0007 & 0.0033 & 0.0007 & 0.0009 \\ 0 & 0 & -0.0019 & 0.0018 \end{bmatrix}. \qquad (19)$$

Zeros appear in the $M_{13}$, $M_{14}$, $M_{41}$, and $M_{42}$ elements. The signs of the $M_{12}$ and $M_{21}$ and $M_{31}$ element are the same (coefficient $b_1$) and the $M_{22}$ element is the same as to the $M_{32}$ element (coefficient $a_2$); whereas, $M_{23}$ is the opposite to $M_{33}$ (coefficient $-a_3$ and $a_3$) and $M_{24}$ and $M_{43}$ are opposite $M_{34}$ (coefficient $-b_2$ and $b_2$). Each of these results is expected and supported by Eq. (16).

## 4. Discussion

'Laboratory observation' scattering experiments have been reported which measured the full Mueller Matrix polarization state of light reflected over a range of scattering angles [13-22]. Models of the scattering process have been made to describe this 'laboratory observation' process [23-25].

Maps of hemispherical scattering models have been extended to include incidence angle variations to more closely approximate the situation in planetary atmospheres [1], and the method is followed in this paper.

Rakovic et al. [24, 25] presented equations for the reflected light from a slab penetrated by a laser beam, and calculated $I_{bs}(\phi, \rho)$, where $\rho$ is the radial distance from where the incident beam initially penetrated a slab. Her model is different from the hemispherical reflection models of perpendicular light scattering followed in this paper. The geometry used by Ben David [26] and Yang et al. [27] is similar to that of Rakovic et al. As seen in Fig. 1 of Yang et al. [27], single scattering is calculated from a point, but projected onto a plane rather than onto a hemisphere, as in this paper. In another approach, Li et al. [20] used an 'inline' model where a single particle is placed between source and detector and the FDTD method is used to derive the transfer Mueller Matrix.

The four different methods reported in the literature [1, 20, 25, 27] are incompatible; their results cannot be directly verified by observation. Similarities and differences between the hemispherical symmetry relations and the symmetries of Rakovic et al. [24, 25] need to be further studied.

**5. Conclusions**

The symmetry relationships explicitly identified in our Eqs. (10) through (16) can be used as a straightforward accuracy check of Mueller matrices for future modeling and experimental work.

For both zenith and off-zenith illumination and for all hexagonal and sphere targets with a plane of symmetry, the symmetry relations may be summarized as:

- Mirror symmetry within elements. The top left block and bottom right block are anti-symmetrical in the x and y planes, and the bottom left and top right are symmetrical in those same planes.
- Circles and crosses. Circles around the incidence beam for elements $M_{11}$ and $M_{44}$. Crosses around the incidence beam for elements $M_{*2}$ and $M_{*3}$ (two central columns). These crosses are offset for off-zenith illumination configurations.
- 45° rotation relationship. $M_{12}(\theta+45°)=M_{13}(\theta)$, $M_{22}(\theta+45°) \rightarrow M_{23}(\theta)$, $M_{32}(\theta+45°) \rightarrow M_{33}(\theta)$, and $M_{42}(\theta+45°) \rightarrow M_{43}(\theta)$. For elements in columns $M_{*2}$ to $M_{*3}$, each element has rotated counterclockwise when looking in the direction of propagation.

For both zenith and off-zenith illumination of spherical targets, the symmetry relations may be summarized as:

- Edge elements symmetries. $M_{21}$ is a linear inverse of $M_{34}$, $M_{42}$ is a linear inverse of $M_{13}$, and $M_{12} \propto M_{43}$. $M_{41} \propto M_{14}$, and the elements tend to approach zero with decreasing optical depth.

For zenith illumination of all targets,

- Null elements. The 1x2 block elements of the Mueller Matrix maps in top right and bottom left (i.e. $M_{14}$, $M_{24}$, $M_{31}$, and $M_{41}$) are always null.

## 6. Acknowledgements

We would like to thank Joop Hovenier and Ping Yang and an anonymous reviewer for their helpful comments on this manuscript. The adding-doubling code was generously provided by Johan de Haan.